\documentclass[prd,twocolumn,showpacs,tightenlines,%
superscriptaddress,nofootinbib,notitlepage,preprintnumbers]{revtex4}

\usepackage{hyperref}
\usepackage{amsmath}
\usepackage{amssymb}

\begin{document}

\title{Stability of a black hole and the speed of gravity waves\\
within self-tuning cosmological models}

\author{Eugeny~Babichev} 
\affiliation{Laboratoire de Physique Th\'eorique,
CNRS, Universit\'e Paris-Sud,
Universit\'e Paris-Saclay,
F-91405 Orsay Cedex, France}
\affiliation{UPMC-CNRS, UMR7095,
Institut d'Astrophysique de Paris,
${\mathcal{G}}{\mathbb{R}}\varepsilon{\mathbb{C}}{\mathcal{O}}$,
98bis boulevard Arago, F-75014 Paris, France}

\author{Christos~Charmousis} 
\affiliation{Laboratoire de Physique Th\'eorique,
CNRS, Universit\'e Paris-Sud,
Universit\'e Paris-Saclay,
F-91405 Orsay Cedex, France}

\author{Gilles~\surname{Esposito-Far\`ese}} 
\affiliation{UPMC-CNRS, UMR7095,
Institut d'Astrophysique de Paris,
${\mathcal{G}}{\mathbb{R}}\varepsilon{\mathbb{C}}{\mathcal{O}}$,
98bis boulevard Arago, F-75014 Paris, France}

\author{Antoine~Leh\'ebel} 
\affiliation{Laboratoire de Physique Th\'eorique,
CNRS, Universit\'e Paris-Sud,
Universit\'e Paris-Saclay,
F-91405 Orsay Cedex, France}


\begin{abstract}
The gravitational wave event GW170817 together with its
electromagnetic counterparts constrains the speed of gravity
to be extremely close to that of light. We first show, on the
example of an exact Schwarzschild-de Sitter solution of a
specific beyond-Horndeski theory, that imposing the strict
equality of these speeds in the asymptotic homogeneous Universe
suffices to guarantee so even in the vicinity of the black hole,
where large curvature and scalar-field gradients are present. We
also find that the solution is stable in a range of the model
parameters. We finally show that an infinite class of
beyond-Horndeski models satisfying the equality of gravity and
light speeds still provide an elegant self-tuning: The very large
bare cosmological constant entering the Lagrangian is almost
perfectly counterbalanced by the energy-momentum tensor of the
scalar field, yielding a tiny observable effective cosmological
constant.
\end{abstract}

\date{December 15, 2017}

\maketitle

The simultaneous detection of the gravitational wave event
GW170817 and its electromagnetic counterparts
\cite{TheLIGOScientific:2017qsa,Monitor:2017mdv} constrains the
speeds of light and gravity to differ by no more than a few parts
in $10^{15}$.
References~\cite{Creminelli:2017sry,Ezquiaga:2017ekz}
(see also \cite{Baker:2017hug}) have characterized which
Horndeski theories
\cite{Horndeski:1974wa,Fairlie,Nicolis:2008in,Deffayet:2009wt,%
Deffayet:2009mn,
Deffayet:2011gz,Deffayet:2013lga},
and their generalizations
\cite{Bettoni:2013diz,Zumalacarregui:2013pma,Gleyzes:2014dya,%
Lin:2014jga,Gleyzes:2014qga,Deffayet:2015qwa,Langlois:2015cwa,%
Langlois:2015skt,Crisostomi:2016tcp,Crisostomi:2016czh,%
Achour:2016rkg,deRham:2016wji,BenAchour:2016fzp},
satisfy exactly $c_\text{grav} = c_\text{light}$ in a homogeneous
Universe. The aim of the present work is twofold. First, we check
that this speed equality remains satisfied even in a very
inhomogeneous situation, namely in the vicinity of a black hole,
where gradients are large and where the separation of spin-2 and
spin-0 degrees of freedom is difficult. This will be done for an
exact Schwarzschild-de Sitter solution of a specific model
\cite{Babichev:2013cya} (see also
\cite{Kobayashi:2014eva,Babichev:2016rlq}). We also report that
this solution is ghost free and has no gradient instability for
some ranges of the parameters defining the theory. We will refer
to such solutions as being \textit{stable} throughout this
letter. We then show that self-tuning cosmological models
\cite{Charmousis:2011bf,Charmousis:2011ea,Babichev:2015qma,%
Appleby:2012rx,Linder:2013zoa,Martin-Moruno:2015bda,%
Martin-Moruno:2015lha,Starobinsky:2016kua,Babichev:2016kdt}
still exist while taking into account the $c_\text{grav} =
c_\text{light}$ constraint. In such models, the energy-momentum
tensor of the scalar field almost perfectly counterbalances the
very large bare cosmological constant assumed to be present in
the Lagrangian, so that the observable accelerated expansion of
the Universe is consistent with a tiny effective cosmological
constant.

In the present paper, we focus on the subclass of shift-symmetric
beyond-Horndeski theories, i.e., which do not involve any
undifferentiated scalar field $\varphi$. Their Lagrangian
reads\footnote{We use the sign conventions of
Ref.~\cite{Misner:1974qy}, and notably the mostly-plus signature.}
\pagebreak
\begin{eqnarray}
\mathcal{L} &=& G_2(\varphi_\lambda^2) + G_3(\varphi_\lambda^2)\,
\Box\varphi +G_4(\varphi_\lambda^2) R\nonumber\\
&&- 2 G_4'\left(\varphi_\lambda^2\right)
\Bigl[\left(\Box\varphi\right)^2
- \varphi_{\mu\nu}\varphi^{\mu\nu}\Bigr]\nonumber\\
&&+F_4\left(\varphi_\lambda^2\right) \varepsilon^{\mu\nu\rho\sigma}\,
\varepsilon^{\alpha\beta\gamma}_{\hphantom{\alpha\beta\gamma}\sigma}\,
\varphi_\mu\,
\varphi_\alpha\, \varphi_{\nu\beta}\, \varphi_{\rho\gamma}\nonumber\\
&&+\mathcal{L}_\text{matter}\left[\psi, g_{\mu\nu}\right],
\label{EqLG}
\end{eqnarray}
where $\varepsilon^{\mu\nu\rho\sigma}$ denotes the
fully-antisymmetric Levi-Civita tensor, $\varphi_\mu \equiv
\partial_\mu \varphi$ and $\varphi_{\mu\nu} \equiv
\nabla_\mu\nabla_\nu \varphi$ are the first and second covariant
derivatives of the scalar field, $G_2$, $G_3$, $G_4$ and $F_4$
are functions of the standard scalar kinetic term\footnote{We do
not denote $\varphi_\lambda^2$ as $X$, because this letter will
be used for a dimensionless variant in Eq.~(\ref{EqX}) below.
} $\varphi_\lambda^2 = g ^{\mu\nu} \varphi_\mu
\varphi_\nu$, $G_4'$ is the derivative of $G_4$ with respect to
its argument ($\varphi_\lambda^2$), and $\psi$ denotes globally
all matter fields (including gauge bosons), assumed to be
minimally coupled to the metric $g_{\mu\nu}$. In order to ensure
$c_\text{grav} = c_\text{light}$, no quintic beyond-Horndeski
term is allowed in this shift-symmetric subclass, and the
function $F_4$ must be related to $G_4$ by
\cite{Creminelli:2017sry,Ezquiaga:2017ekz}\footnote{Note
that there is a sign mistake in Eq.~(11) of
Ref.~\cite{Creminelli:2017sry}, and that
Ref.~\cite{Ezquiaga:2017ekz} defines $F_4$ with an opposite sign.}
\begin{equation}
F_4(\varphi_\lambda^2) = -\frac{2\,
G_4'(\varphi_\lambda^2)}{\varphi_\lambda^2}.
\label{EqBigF4}
\end{equation}
This condition (\ref{EqBigF4}) ensures that the speeds of
gravitational and electromagnetic waves coincide at least in a
homogeneous cosmological background. However the waves of the
GW170817 event did pass nearby massive bodies during their 40~Mpc
journey, and if their speeds slightly differed in such
inhomogeneous situations, this would \textit{a priori} suffice to
increase the delay between their detections. It is thus important
to check that these speeds remain equal even in inhomogeneous
backgrounds. Actually, Ref.~\cite{Ezquiaga:2017ekz} claims that
condition (\ref{EqBigF4}) also suffices around arbitrary
backgrounds, and we will confirm so below for a specific exact
solution ---~see Eq.~(\ref{EqGmunueff}).
But this reference \cite{Ezquiaga:2017ekz} uses the
results of \cite{Bettoni:2016mij}, which needed to neglect
scalar-graviton mixing terms in order to extract the spin-2
excitations. Generically, the separation of the spin-2 and spin-0
degrees of freedom is background dependent and highly
non-trivial. The same argument applies to the ADM decomposition
of the Lagrangian (\ref{EqLG}) under condition (\ref{EqBigF4}).
In the unitary gauge, one recovers the same decomposition as in
general relativity \cite{Langlois:2017dyl,Creminelli:2017sry}.
However, this does not necessarily mean that the helicity-2
perturbations propagate at $c_\text{grav}=c_\text{light}$,
because the Lagrangian contains mixing terms proportional to
$\dot{h}_{ij}D_iN_j$, in usual ADM notation; the shift $N_i$
cannot be eliminated as in general relativity, because the gauge
is already fixed. It remains thus important to check whether this
speed equality is also satisfied in very curved backgrounds, with
large scalar-field gradients, for instance in the neighborhood of
a black-hole horizon. This is what we do now for the particular
case of an exact Schwarzschild-de Sitter solution of the model
\begin{equation}
\mathcal{L} = \zeta\left(R-2\Lambda_\text{bare}\right)
-\eta\, \varphi_\lambda^2
+ \beta\, G^{\mu\nu}\varphi_\mu\varphi_\nu,
\label{EqJohn}
\end{equation}
where $G^{\mu\nu}$ denotes the Einstein tensor, $\zeta =
\frac{1}{2}M_\text{Pl}^2 > 0$, and $M_\text{Pl} \equiv
(8\pi G)^{-1/2}$ is the reduced Planck mass. (The
$G^{\mu\nu}\varphi_\mu\varphi_\nu$ term has been nicknamed
``John'' in the ``Fab-Four'' model
\cite{Charmousis:2011bf,Charmousis:2011ea}.) In terms of the
notation of Eq.~(\ref{EqLG}), this corresponds to
\begin{eqnarray}
G_2(\varphi_\lambda^2) &=& - 2\,\zeta\, \Lambda_\text{bare}
- \eta\, \varphi_\lambda^2,
\label{EqG2J}\\
G_4(\varphi_\lambda^2) &=& \zeta
- \frac{\beta}{2}\, \varphi_\lambda^2,
\label{EqG4J}
\end{eqnarray}
and $G_3=F_4=0$. Since this vanishing of $F_4$ is in
contradiction with Eqs.~(\ref{EqBigF4}) and (\ref{EqG4J}), we can
immediately conclude that this model does not satisfy the
$c_\text{grav} = c_\text{light}$ constraint, if matter (and
thereby light) is assumed to be minimally coupled to $g_{\mu\nu}$
as in Eq.~(\ref{EqLG}). However, as already underlined in
\cite{Creminelli:2017sry,Ezquiaga:2017ekz}, it suffices to couple
matter to a different metric $\tilde g_{\mu\nu}$, related to
$g_{\mu\nu}$ by a disformal transformation, to change the matter
causal cone so that $c_\text{grav} = c_\text{light}$ is ensured,
at least in a homogeneous Universe. In the present model, the
disformal transformations given in
\cite{
Bettoni:2013diz,Zumalacarregui:2013pma,%
Crisostomi:2016czh,Achour:2016rkg} or the gravity speed derived
in \cite{Bettoni:2016mij,Ezquiaga:2017ekz} allow us to prove that
this physical metric must read (or be proportional to)
\begin{equation}
\tilde g_{\mu\nu} = g_{\mu\nu} - \frac{\beta}{\zeta+\frac{\beta}{2}\,
\varphi_\lambda^2}\, \varphi_\mu \varphi_\nu.
\label{Eqgtilde}
\end{equation}
One may also rewrite Lagrangian (\ref{EqJohn}) in terms of this
$\tilde g_{\mu\nu}$, and one finds that it becomes of the form
(\ref{EqLG}), with rather complicated functions $\tilde
G_4(\tilde \varphi_\lambda^2)$ and $\tilde F_4(\tilde
\varphi_\lambda^2)$ (involving nested square roots), which now do
satisfy the constraint~(\ref{EqBigF4}) in terms of the variable
$\tilde \varphi_\lambda^2 \equiv \tilde g^{\mu\nu}\varphi_\mu
\varphi_\nu$. This guarantees that the speeds of light and
gravity coincide at least in the asymptotic homogeneous Universe,
far away from any local massive body.

We can go beyond this result by studying the speed of spin-2
perturbations around a spherical black hole. An exact
Schwarzschild-de Sitter solution has indeed been found in
\cite{Babichev:2013cya} for model (\ref{EqJohn}), assuming linear
time-dependence of the scalar field~\cite{Babichev:2010kj}:
\begin{eqnarray}
\label{Eqds2}
ds^2 &=& -A(r)\, dt^2
+\frac{dr^2}{A(r)}
+ r^2\left(d\theta^2 +\sin^2\theta\, d\phi^2\right),~~\\
A(r) &=& 1- \frac{2Gm}{r} - \frac{\Lambda_\text{eff}}{3}\, r^2,\\
\Lambda_\text{eff} &=& -\frac{\eta}{\beta},
\label{EqLambdaEff}\\
\varphi &=& q\left( t - \int\frac{\sqrt{1-A(r)}}{A(r)} dr\right),
\label{Eqphi}\\
q^2 &=& \frac{\eta+ \beta\, \Lambda_\text{bare}}{\eta\,\beta}\,\zeta,
\label{Eqq2}
\end{eqnarray}
where this last equality (\ref{Eqq2}) forces its right-hand side
to be positive. Equation~(\ref{EqLambdaEff}) defines the
effective cosmological constant $\Lambda_\text{eff}$ entering the
line element (\ref{Eqds2}), and one can note that it may be as
small as one wishes, independently of the magnitude of
$\Lambda_\text{bare}$ (it does not even depend at all on
$\Lambda_\text{bare}$, in the present model). This is a
particularly simple example of self-tuning. However, the observer,
made of matter, is now assumed to be coupled to the physical
metric (\ref{Eqgtilde}), and this changes her perception of the
Universe. A straightforward calculation shows that $\tilde
g_{\mu\nu}$ remains of the exact Schwarzschild-de Sitter form,
with a scalar field of the form (\ref{Eqphi}) in the relevant
transformed coordinate system, but the observable cosmological
constant now reads
\begin{equation}
\tilde \Lambda_\text{eff} = \left(\frac{\Lambda_\text{eff}
+ \Lambda_\text{bare}}{3\Lambda_\text{eff}-\Lambda_\text{bare}}\right)
\Lambda_\text{eff}.
\label{EqLambdaEffTilde}
\end{equation}
At this stage, it thus seems that a very small $\tilde
\Lambda_\text{eff}$ remains possible, for instance if
$\Lambda_\text{eff} = -\eta/\beta$ is chosen to almost compensate
$\Lambda_\text{bare}$. However, the field equations written in
the physical frame $\tilde g_{\mu\nu}$ actually always imply
$\tilde \Lambda_\text{eff} \sim \Lambda_\text{bare}$
\cite{Babichev:2018}. Moreover, we will see below that the
stability of the solution forces both $\Lambda_\text{eff}$ and
the observable $\tilde \Lambda_\text{eff}$ to be of the same
order of magnitude as $\Lambda_\text{bare}$ (or even larger).
Therefore, in this simple model (\ref{EqJohn}), the small
observed cosmological constant cannot be explained by the
self-tuning mechanism, and some other reason must be invoked,
like in standard general relativity. It remains that this model
is observationally consistent if the constant
$\Lambda_\text{bare}$ entering (\ref{EqJohn}) is small enough.

The odd-parity perturbations of solution
(\ref{Eqds2})--(\ref{Eqq2}) have been analyzed in
\cite{Ogawa:2015pea}, and they define the effective metric, say
$\mathcal{G}_{\mu\nu}$, in which spin-2 perturbations propagate.
We can thus compare it with the metric $\tilde g_{\mu\nu}$,
Eq.~(\ref{Eqgtilde}), to which matter (including photons) is
assumed to be coupled, and we find
\begin{eqnarray}
\mathcal{G}_{\mu\nu} =
\left(\frac{\Lambda_\text{eff}}{\Lambda_\text{eff}
+\Lambda_\text{bare}}\right)
\tilde{g}_{\mu\nu}.
\label{EqGmunueff}
\end{eqnarray}
Therefore, even close to the black hole, their causal cones
exactly coincide. In other words, the universal coupling of
matter to the disformal metric (\ref{Eqgtilde}) suffices to
ensure $c_\text{grav} = c_\text{light}$ even in a very
inhomogeneous configuration. Details will be given in
\cite{Babichev:2018}.

The perturbative analysis of Ref.~\cite{Ogawa:2015pea} was
actually performed to claim that the above Schwarzschild-de
Sitter solution is always unstable, but this claim is incorrect.
The argument was that the Hamiltonian of these perturbations is
unbounded by below, close enough to the black-hole horizon.
However, although a bounded Hamiltonian does guarantee the
stability of the lowest-energy state, the converse theorem does
not exist. Indeed, a Hamiltonian is not a scalar quantity, and it
may take large negative values in a very boosted frame even if
it was bounded by below initially. It gets mixed with other
conserved quantities which are not bounded by below,
corresponding to the 3-momentum of the system. A correct
stability criterion may thus be formulated as: If the Hamiltonian
is bounded by below in at least one coordinate system, then the
solution is stable. As we shall detail in \cite{Babichev:2018},
when focusing on kinetic terms, it suffices that the causal cones
of all interacting degrees of freedom share a common interior
timelike\footnote{Timelike (resp. spacelike) means here that for
an effective metric $\mathcal{G}_{\mu\nu}$ in which a given
degree of freedom propagates, the line element
$\mathcal{G}_{\mu\nu} dx^\mu dx^\nu$ is negative (resp.
positive). Our criterion on the intersections of causal cones
therefore also forbids the existence of ghost degrees of freedom,
for which the effective metric would have the opposite
(mostly-minus) signature.} direction (which will become the time
coordinate of the ``safe'' frame in which the Hamiltonian can be
proven to be bounded by below), and also a common exterior
spacelike hypersurface, on which initial data may be set to
define the Cauchy problem
(see \cite{Babichev:2006vx,Bruneton:2006gf,%
Bruneton:2007si,Babichev:2007dw,Hassan:2017ugh}
for related discussions).

In the present solution (\ref{Eqds2})--(\ref{Eqq2}) for
model (\ref{EqJohn}), we saw that the graviton and matter causal
cones coincide everywhere, because of the disformal metric
(\ref{Eqgtilde}) to which we universally couple matter. There
remains however to check that both interiors are indeed timelike,
otherwise matter and gravitons would have opposite kinetic
energies. This means that the factor entering
Eq.~(\ref{EqGmunueff}) must be positive. Moreover, the scalar
field $\varphi$ itself has a different causal cone, that we study
in \cite{Babichev:2018} by analyzing the $\ell = 0$ even-parity
perturbations. Therefore, stability can be ensured only if the
scalar causal cone shares a common interior direction and a
common exterior hypersurface with that of $\tilde g_{\mu\nu}$
(and gravitons). We found that it is possible provided the
parameters of Lagrangian (\ref{EqJohn}) satisfy the following
inequalities. We write them here in terms of the ratio
$-\eta/\beta$, denoted as $\Lambda_\text{eff}$ in
Eq.~(\ref{EqLambdaEff}), and we assume that the observed
$\tilde\Lambda_\text{eff} = 3\tilde H^2$ is positive (which
implies that $\Lambda_\text{bare}$ and $\Lambda_\text{eff}$ are
positive too, when taking into account these very inequalities):
\begin{eqnarray}
\hbox{either $\eta > 0$, $\beta < 0$,} &\hbox{and}&
\frac{1}{3}\,\Lambda_\text{bare} < -\frac{\eta}{\beta}
< \Lambda_\text{bare},\quad
\label{EqIneq1}\\
\hbox{or $\eta < 0$, $\beta > 0$,} &\hbox{and}&\quad
\Lambda_\text{bare} < -\frac{\eta}{\beta} < 3\Lambda_\text{bare}.
\qquad
\label{EqIneq2}
\end{eqnarray}
It is straightforward to prove analytically that these conditions
suffice for the consistency of the causal cones in the asymptotic
de Sitter Universe. Close to the black hole, the analytic
expressions are so long that we checked instead specific examples
in a numerical way, by following the relative positions of the
scalar and graviton (or matter) causal cones while varying the
distance $r$ to the center of the black hole. Our conclusion is
that for the above ranges of the ratio $-\eta/\beta$,
Eqs.~(\ref{EqIneq1}) or (\ref{EqIneq2}), the perturbation
Hamiltonian is bounded by below in a well-chosen frame, at any
spacetime point, and the stability criterion we established is
satisfied.

Note that when setting $\Lambda_\text{bare}=0$, the interval of
stability disappears, in agreement with the perturbation analysis
of \cite{Appleby:2011aa} around a cosmological background. In
other words, it is the presence of vacuum energy which allows for
a window of stability for the black hole solution.

In terms of the observed $\tilde\Lambda_\text{eff}$,
Eq.~(\ref{EqLambdaEffTilde}), conditions (\ref{EqIneq1}) and
(\ref{EqIneq2}) imply
\begin{eqnarray}
\hbox{either $\eta > 0$, $\beta < 0$,}
&\hbox{and}&\Lambda_\text{bare} < \tilde\Lambda_\text{eff},\\
\hbox{or $\eta < 0$, $\beta > 0$,}
&\hbox{and}&\Lambda_\text{bare} < \tilde\Lambda_\text{eff} <
\frac{3}{2}\,\Lambda_\text{bare}.\qquad
\end{eqnarray}
As stressed below Eq.~(\ref{EqLambdaEffTilde}), this means that
self-tuning is impossible in this specific model, since the
observed cosmological constant must always be larger than the
bare one.

But there exists an infinite class of other beyond-Horndeski
models which do provide self-tuning, as shown in
Ref.~\cite{Babichev:2016kdt}, and we prove below that a subclass
of them also satisfies the $c_\text{grav} = c_\text{light}$
constraint. From now on, we assume that matter is minimally
coupled to $g_{\mu\nu}$, as in Lagrangian (\ref{EqLG}), and we no
longer consider any disformal transformation such as
(\ref{Eqgtilde}).

To avoid hiding several different scales in the functions of
$\varphi_\lambda^2$, it is convenient to work with the
dimensionless quantity
\begin{equation}
X \equiv \frac{-\varphi_\lambda^2}{M^2},
\label{EqX}
\end{equation}
$M$ being the only mass scale entering the Lagrangian of the
scalar field $\varphi$, itself chosen dimensionless (beware not
to confuse $M$ with the Planck mass $M_\text{Pl}$). All the
coefficients entering dimensionless functions of $X$ will also
be assumed to be of order $\mathcal{O}(1)$. Up to a total
derivative, action (\ref{EqLG}) may then be rewritten
as\footnote{The above model (\ref{EqJohn}) corresponds for
instance to constant values $\zeta = \frac{1}{2}M_\text{Pl}^2$,
$f_2 = -\eta/M^2$, $s_4 = -\frac{1}{4}\beta$, and $f_4 = 0$.}
\begin{eqnarray}
\mathcal{L}&\!\!=\!\!&\frac{M_\text{Pl}^2}{2}
\left(R-2\Lambda_\text{bare}\right)
-M^4 X f_2(X)
- 4 s_4(X) G^{\mu\nu} \varphi_\mu \varphi_\nu\nonumber\\
&&-\frac{f_4(X)}{M^2}\, \varepsilon^{\mu\nu\rho\sigma}\,
\varepsilon^{\alpha\beta\gamma}_{\hphantom{\alpha\beta\gamma}\sigma}\,
\varphi_\mu\, \varphi_\alpha\,
\varphi_{\nu\beta}\, \varphi_{\rho\gamma}
+\mathcal{L}_\text{matter},
\label{EqLf}
\end{eqnarray}
where we do not include the $G_3$ term because it must anyway be
passive for the self-tuning solutions derived in
\cite{Babichev:2016kdt} (see this reference for the explicit
translation between (\ref{EqLG}) and (\ref{EqLf}) as well as
other notation used in the literature).

The $c_\text{grav} = c_\text{light}$ constraint (\ref{EqBigF4})
becomes then
\begin{equation}
f_4(X) = -\frac{4s_4(X)}{X} ,
\label{EqSmallf4}
\end{equation}
while $f_2(X)$ and $s_4(X)$ are arbitrary. For monomials,
this means that we need
\begin{equation}
f_2 = k_2 X^\alpha,\quad s_4 = \kappa_4 X^\gamma,\quad
f_4 = -4\kappa_4 X^{\gamma-1}.
\label{Eqf4}
\end{equation}
where $k_2$, $\kappa_4$, $\alpha$ and $\gamma$ are dimensionless
constants of order $\mathcal{O}(1)$. Note that negative exponents
$\alpha$ and $\gamma$ are perfectly allowed and consistent in
this cosmological context, where the background solution
corresponds to a strictly positive value of $X$. Perturbations
are thus well-defined around such a background.

Particular self-tuning models respecting $c_\text{grav} =
c_\text{light}$ are thus easily obtained from~(\ref{Eqf4}).
However, it should be stressed that it is not enough to find a
theory with $\Lambda_\text{eff} \ll \Lambda_\text{bare}$, since
Newton's constant also generically gets renormalized, giving
$\left(M_\text{Pl}^2 \Lambda\right)_\text{eff} \sim
\left(M_\text{Pl}^2
\Lambda\right)_\text{bare}$~\cite{Babichev:2016kdt}. But luckily
a subclass of models (\ref{Eqf4}) is such that $M_\text{Pl}$
remains unrenormalized, and it is thus possible to get
$M_\text{Pl}^2 \Lambda_\text{eff} \ll M_\text{Pl}^2
\Lambda_\text{bare}$ by choosing an appropriate value of $M$.
This subclass corresponds to the exponent $\gamma =
-\frac{3}{2}$, i.e., $s_4 = \kappa_4 X^{-3/2}$ and $f_4 =
-8\kappa_4 X^{-5/2}$. In terms of the $G_i$ notation of
Eq.~(\ref{EqLG}), this reads
\begin{eqnarray}
G_2(\varphi_\lambda^2) &=& - M_\text{Pl}^2 \Lambda_\text{bare}
- k_2 M^4 \left(\frac{-\varphi_\lambda^2}{M^2}\right)^{\alpha+1},
\label{EqG2}\\
G_4(\varphi_\lambda^2) &=& \frac{1}{2} M_\text{Pl}^2
- 2 \kappa_4 M^3 \left(-\varphi_\lambda^2\right)^{-1/2},
\label{EqG4b}
\end{eqnarray}
while $F_4$ is given by Eq.~(\ref{EqBigF4}).
One then finds that the Schwarzschild-de Sitter equations of
Ref.~\cite{Babichev:2016kdt} can be solved provided $\alpha \neq
-1$ and $\alpha\neq -\frac{1}{2}$, and they imply
\begin{equation}
(H^2)^{\alpha+1}(M^2)^{\alpha+2} \propto
(M_\text{Pl}^2 \Lambda_\text{bare})^{\alpha+3/2}.
\label{EqMagnitude}
\end{equation}
The proportionality factor depends on the
$\mathcal{O}(1)$-dimensionless constants $k_2$, $\kappa_4$ and
$\alpha$, and is thus itself of order $\mathcal{O}(1)$.
Therefore, if $\alpha\neq -2$, it suffices to choose $M$
appropriately to get $H$ equal to the observed value, whatever
the large $\Lambda_\text{bare}$ entering the action. Note that
\textit{all} these models (with $\alpha \not\in\{-2, -1,
-\frac{1}{2}\}$ and $\gamma = -\frac{3}{2}$) do admit exact
Schwarzschild-de Sitter solutions such that $M_\text{Pl}^2
\Lambda_\text{eff}$ is consistent with its small observed value,
and they also satisfy $c_\text{grav} = c_\text{light}$ at least
in the asymptotic homogeneous Universe.

As underlined at the very end of \cite{Babichev:2016kdt}, if
the bare cosmological constant happens to take the huge value
$\Lambda_\text{bare} \sim M_\text{Pl}^2$, then the particular
case $\alpha=-\frac{5}{4}$ needs a rather natural value of the
scale $M \sim 100~\text{MeV}$, similar to usual elementary
particle masses.

Another interesting particular case is $\alpha=-\frac{3}{2}$, for
instance $f_2 = s_4 = -X^{-3/2}$ and $f_4 = 4 X^{-5/2}$ (choosing
here $k_2 = \kappa_4 = -1$ to simplify, the signs being imposed
by the field equations). This corresponds to
\begin{eqnarray}
G_2(\varphi_\lambda^2) &=& - M_\text{Pl}^2 \Lambda_\text{bare}
+ M^5 \left(-\varphi_\lambda^2\right)^{-1/2},
\label{EqG2c}\\
G_4(\varphi_\lambda^2) &=& \frac{1}{2} M_\text{Pl}^2
+ 2 M^3 \left(-\varphi_\lambda^2\right)^{-1/2},
\label{EqG4c}\\
F_4(\varphi_\lambda^2) &=&
2M^3 \left(-\varphi_\lambda^2\right)^{-5/2}.
\label{EqF4c}
\end{eqnarray}
Then the exact version of Eq.~(\ref{EqMagnitude}) implies that
one must choose $M = 2\sqrt{3}\, H$, i.e., the very small
observed Hubble expansion rate $H$ must actually be put by hand
in the action \textit{via} the scale $M$. But this drawback comes
with the great bonus that this observed $H$ now depends only on
$M$, and no longer on the bare vacuum energy density
$M_\text{Pl}^2 \Lambda_\text{bare}$. Therefore, even if
$\Lambda_\text{bare}$ happens to change because of a phase
transition during the cosmological evolution of the Universe, the
effective $\Lambda_\text{eff} = (M/2)^2 = 3 H^2$ remains constant
and small.

The important conclusion is that elegant self-tuning cosmological
models are still allowed, even when taking into account the
experimental constraint $c_\text{grav} = c_\text{light}$. Note
that for these models, we did not prove that the speed equality
remains valid in the vicinity of massive bodies. However, our
result above for the simple model (\ref{EqJohn}) and the argument of
Refs.~\cite{Creminelli:2017sry,Ezquiaga:2017ekz,Langlois:2017dyl}
show that it may remain true, at least for Schwarzschild-de
Sitter black-hole solutions. The stability of these self-tuning
models should also be analyzed, as we did above for model
(\ref{EqJohn}). Aside of this, it would be of great interest to
study a more realistic cosmological evolution for these
self-tuning models, as in \cite{Starobinsky:2016kua}, where
certain branches of solutions were shown to screen matter as well
as the cosmological constant.

\acknowledgments{We thank G.~Bougas for collaboration and
discussions in the early stages of this project. We thank
M.~Crisostomi, A.~Fabbri and K.~Noui for many interesting
discussions. We acknowledge support from the French research
programs ``Programme national de cosmologie et galaxies'' of
the CNRS/INSU and from the call ``D\'efi InFIniti'' 2017.}


\end{document}